\pdfoutput=1
\documentclass[prd,aps,twocolumn,preprintnumbers,amsmath,amssymb,superscriptaddress]{revtex4}

\usepackage{graphicx}
\usepackage{dcolumn}
\usepackage{bm}
\usepackage{color}

\begin{document}


\title{Systematic uncertainties in the determination of the local dark matter density}

\author{Miguel Pato}
\email{pato@iap.fr}
\affiliation{Institute for Theoretical Physics, Univ.~of Z\"urich, Winterthurerst. 190, 8057 Z\"urich CH}
\affiliation{Institut d'Astrophysique de Paris, UMR 7095-CNRS, Univ.~Pierre \& Marie Curie, 98bis Bd Arago 75014 Paris, France}
\affiliation{Dipartimento di Fisica, Universit\`a degli Studi di Padova, via Marzolo 8, I-35131, Padova, Italy}

\author{Oscar Agertz}
\affiliation{Institute for Theoretical Physics, Univ.~of Z\"urich, Winterthurerst. 190, 8057 Z\"urich CH}

\author{Gianfranco Bertone}
\affiliation{Institute for Theoretical Physics, Univ.~of Z\"urich, Winterthurerst. 190, 8057 Z\"urich CH}
\affiliation{Institut d'Astrophysique de Paris, UMR 7095-CNRS, Univ.~Pierre \& Marie Curie, 98bis Bd Arago 75014 Paris, France}

\author{Ben Moore}
\affiliation{Institute for Theoretical Physics, Univ.~of Z\"urich, Winterthurerst. 190, 8057 Z\"urich CH}

\author{Romain Teyssier}
\affiliation{Institute for Theoretical Physics, Univ.~of Z\"urich, Winterthurerst. 190, 8057 Z\"urich CH}

\date{\today}

\begin{abstract}
A precise determination of the local dark matter density and an accurate control over the corresponding uncertainties are of paramount importance for Dark Matter (DM) searches. Using very recent high-resolution numerical simulations of a Milky Way like object, we study the systematic uncertainties that affect the determination of the local dark matter density based on dynamical measurements in the Galaxy.  In particular, extracting from the simulation with baryons the orientation of the Galactic stellar disk with respect to the DM distribution, we study the DM density for an observer located at $\sim$8 kpc from the Galactic center {\it on the stellar disk}, $\rho_0$. This quantity is found to be always larger than the average density in a spherical shell of same radius $\bar{\rho}_0$, which is the quantity inferred from dynamical measurements in the 
Galaxy, and to vary in the range $\rho_0/\bar{\rho}_0=1.01-1.41$. This suggests that the actual dark matter density in the solar neighbourhood is on average 21\% larger than the value inferred from most dynamical measurements, and that the associated systematic errors are larger than the statistical errors recently discussed in the literature. 
\end{abstract}

\keywords{Suggested keywords}
\maketitle


\section{Introduction}\label{secintro}

A wide array of experimental strategies have been devised in order to identify the nature of dark matter (DM) \cite{Bergstrom:2000pn,Munoz:2003gx,Bertone:2004pz,book}. A key parameter in many of these searches is the local density of DM, namely the density of DM particles in the solar neighbourhood, $\rho_0$. For instance, the rate of events in direct detection experiments, that seek to measure the recoil energy in scattering events of DM particles off nuclei in the detector, is obviously proportional to the flux of DM particles through the detector, which in turn is directly proportional to the local DM density. Similarly, the neutrino flux from DM annihilation in the Sun is proportional to the capture rate of DM particles, in turn proportional to the flux of DM particles through the Sun, and therefore to  $\rho_0$. As for indirect searches, the predicted flux of secondary particles, produced by the annihilation of DM particles, is proportional to  $\rho_0^2$. A careful determination of this quantity is therefore of paramount importance in order to extract the properties of DM particles, especially when trying to perform a combined analysis of direct detection and LHC data ~\cite{LHCdirect}.

Interestingly, as also pointed out in Ref.~\cite{Salucci}, $\rho_0$ is often assumed to be equal to 0.3 GeV cm$^{-3}$ , with an error of {\it a factor of 2} . However, this value is often given without a reference, and when a reference is given, it can be traced back to papers which are a few decades old, e.g.~\cite{CO,GGT}. Notice that the link between observed galaxy masses and their host dark matter halo masses has been studied extensively using satellite kinematics \cite{Conroy,More}, weak lensing \cite{Mandelbaum} as well as via abundance matching \cite{Guo}. Due to the large spread in halo characteristics (mass and concentration) at a given galactic baryon mass, these studies are not suitable for inferring detailed properties of our own Milky Way (MW) galaxy, e.g.~the local dark matter density at the solar radius. While the virial mass of the Milky Way is not tightly constrained (see discussion in section \ref{sec2a}), more can be said about the local dark matter density via mass modeling using observational constraints of the galactic baryons and halo stars e.g.~\cite{Klypin}. 

A number of papers have appeared recently on this subject, where the authors attack the problem of the determination of $\rho_0$ in light of recent observational results \cite{Salucci,CatenaUllio,StrigariTrotta}. In Ref.~\cite{CatenaUllio} (see also \cite{StrigariTrotta}), for instance, the authors considered a large set of observational constraints of the Milky Way galaxy, e.g.~the local stellar surface density, the angular rest-frame velocity at the solar radius and inner dark matter halo mass estimates from the velocity dispersion of halo stars. By adopting a Bayesian approach to mass modelling of the Milky Way components, a local dark matter density of $0.385\pm0.027 {\textrm{ GeV}}/\textrm{cm}^{3}$ (assuming an Einasto profile) was found. The quoted $1\sigma$ errors are smaller than in previous studies due to the large set of input constraints, as well as the tight range that exists on a few of them (e.g.~the combination of Oort's constants $A-B$). This value was found to be quite insensitive to the assumed spherical DM density profile. As we shall demonstrate, the non-sphericity of the dark halo and its reaction to galaxy formation introduces larger systematic uncertainties. An alternative technique has been proposed in Ref.~\cite{Salucci}, where a constraint on $\rho_0$ is obtained based on local observables and with presumably no dependence on the mass model of our Galaxy.

Here we estimate the systematic uncertainty on $\rho_0$, with specific emphasis on the impact of departures from spherical symmetry. This has been studied in Ref.~\cite{ghalo} using pure dark matter numerical simulations. We focus instead on a high-resolution simulation of a Milky Way like galaxy \cite{ATM10} $-$ which reproduces the correct properties of our Galaxy $-$ and consider its realizations with and without baryons. In particular, extracting from the simulation with baryons the orientation of the Galactic stellar disk with respect to the DM distribution, we study the DM density for an observer located at $\sim$8 kpc from the Galactic center {\it on the stellar disk}, and show that it is  {\it systematically larger} than the average density in a spherical shell of the same radius. The latter is the observable that has usually been inferred from dynamical constraints such as the local circular velocity, terminal velocities and velocity dispersions of tracer star populations. Notice that we are disregarding fine grained structures (such as microhalos and streams) since their effect on DM searches is likely negligible as shown in \cite{micro}.

The paper is organised as follows: we introduce in section \ref{sims} the numerical simulations of Milky Way sized halos with and without baryons. In Section \ref{syst} we discuss the shape and enclosed mass of the DM component, and the impact on the determination of the local DM density from dynamical measurements. We then conclude in Section \ref{conc}.

\section{Cosmological simulations of galaxy formation}
\label{sims}

\subsection{Models for the Milky Way halo}\label{sec2a}

In order to design a numerical model for the Milky Way halo, we first need to estimate our target halo mass and merging history. We then can explore
large scale $N$-body simulations and pick a good Milky Way candidate, based on our a priori knowledge of the Milky Way halo. In this respect, it is usually admitted
that a good Milky Way candidate should have its total mass between, say, $5 \times 10^{11}$~M$_{\odot}$ and $2 \times 10^{12}$~M$_\odot$ \cite{Xue,Guo}, 
where the mass of a halo is 
defined here as $M_{200c}$, the mass enclosed in a sphere of radius $R_{200c}$ in which the average density is 200 times the critical density at that redshift. Although this 
mass range is relatively small (a factor of 4), choosing a large halo mass, say above $10^{12}$~M$_{\odot}$, rather than a low halo mass, say below $10^{12}$~M$_{\odot}$, has profound consequences on the baryons and dark matter dynamics and their final properties. This is due to the role played by the baryons inside the dark matter halo, and the fact that baryons dissipate their thermal energy, leading to the formation of dense and concentrated objects, usually in the form of centrifugally supported disks \cite{White}. Moreover, baryons are much better
constrained by observations than dark matter. For the Milky Way, we know that the disk can be decomposed in a pseudo-bulge of stellar mass $M_{\rm B} \simeq 1-2 \times 10^{10}$~M$_{\odot}$ and 
a rather extended disk of mass $M_{\rm D} \simeq 5-6 \times 10^{10}$~M$_{\odot}$ \cite{Klypin,Sofue}. We also know that the Milky Way rotation curve is rather flat, with a maximum measured velocity of 220-250~km/s (\cite{Dehnen} and references therein). For large halo mass models, this requires that the baryon fraction within the halo is a factor of 2 below the universal fraction ($\Omega_b/\Omega_m \simeq 0.17$), leading to the so--called ``missing baryons problem" and requiring considerable amount of feedback to eject these baryons out of the halo boundaries (\cite{Anderson} and references therein). The second important consequence is that in these models, the halo is dark matter dominated. The dark matter distribution will be therefore only mildly affected by the baryons dynamics and pure N-body simulations can be reliably used to estimate the local DM density at the Sun radius \cite{Pedrosa}. For small halo mass models, on the other hand, the baryon fraction is close to the universal value. In this case, the effects of baryons dynamics are maximal, with strong adiabatic contraction of the dark halo leading to a change of the global halo shape and concentration, as well as the formation of dark disks \cite{Debattista,Read,Read2,Abadi}. In this second scenario, in order to realistically estimate the dark matter distribution, we need to model the baryonic physics and the formation of the stellar disk. We now describe such a model in the next section. 

\subsection{Baryons dominated Milky Way model}

Agertz, Teyssier \& Moore \cite{ATM10} (from now on ATM10) performed a large suite of cosmological simulations aimed to study the assembly of Milky Way like galaxies. The simulations were carried out using the Adaptive Mesh Refinement (AMR) code {\tt RAMSES}, which includes treatment of dark matter, gas and stars. The gas dynamics is calculated using a second-order unsplit Godunov method, while collisionless particles (including stars) are evolved using the Particle-Mesh technique. The modelling includes realistic recipes for star formation \cite{Rasera}, supernova feedback (SNII and SNIa), stellar mass loss, gas cooling/heating and metal enrichment \cite{Dubois}.

ATM10 demonstrated that realistic disk galaxies form if the efficiency of star formation is low at high redshift, due to small scale processes. We analyze the dark matter component in two of their hydro + $N$-body simulations: SR6-n01e1ML and SR6-n01e5ML. The former simulation adopts an inefficient star formation law (1\% per gas free fall time) and the latter a more efficient one (5\%). These assumptions result in a strong difference in galactic disk size and concentration, hence dark matter contraction, as the bulge-to-disk ratio increases from $\sim 0.25$ to $1.3$. This represents a shift in galactic Hubble type; the disk in the efficient simulation resembles an S0/Sa galaxy while the inefficient one is closer to a Sb/Sbc. These simulations were shown to bracket the observed Kennicutt-Schmidt relation from the THINGS survey \cite{Bigiel}, and are therefore representative of typical $z=0$ disk galaxies. In addition, we compare the results from these simulations to a pure dark matter realization of the same halo.

\par Note that these numerical simulations have a spatial resolution of approximately 340 pc (for the galaxies under study), which is good enough to resolve the baryonic physics taking place in the stellar disk. Such is not achieved by other recent simulations that present resolutions of a few kpc \cite{Pedrosa,Abadi,Gustafsson,Tissera}. \cite{ling} and \cite{Read2} studied the effect of disk formation on the local dark matter density, specifically the formation of a dark disk component. These studies adopted a similar spatial resolution as in this work (few 100 pc), but the galaxies featured very massive bulge components, making them compare unfavorably to late type galaxies like the Milky Way. Hence, the lack of a massive bulge in our simulations signals a more realistic galaxy assembly, and possibly also a more realistic baryonic effect on the dark matter halo structure.

\section{Systematic uncertainties on the local dark matter density}\label{syst}

Any departure from spherical symmetry and any modification of the DM enclosed mass due to adiabatic contraction is expected to lead to a systematic error in the determination of the local dark matter density at the solar position, as in general it will be different from the average density on a shell of same radius, which is the quantity inferred from most dynamical measurements. In this section we study these two effects and quantify the corresponding systematic errors on the determination of $\rho_0$.

\begin{figure}[t]
 \centering
 \includegraphics[width=8.5cm,height=6.5cm]{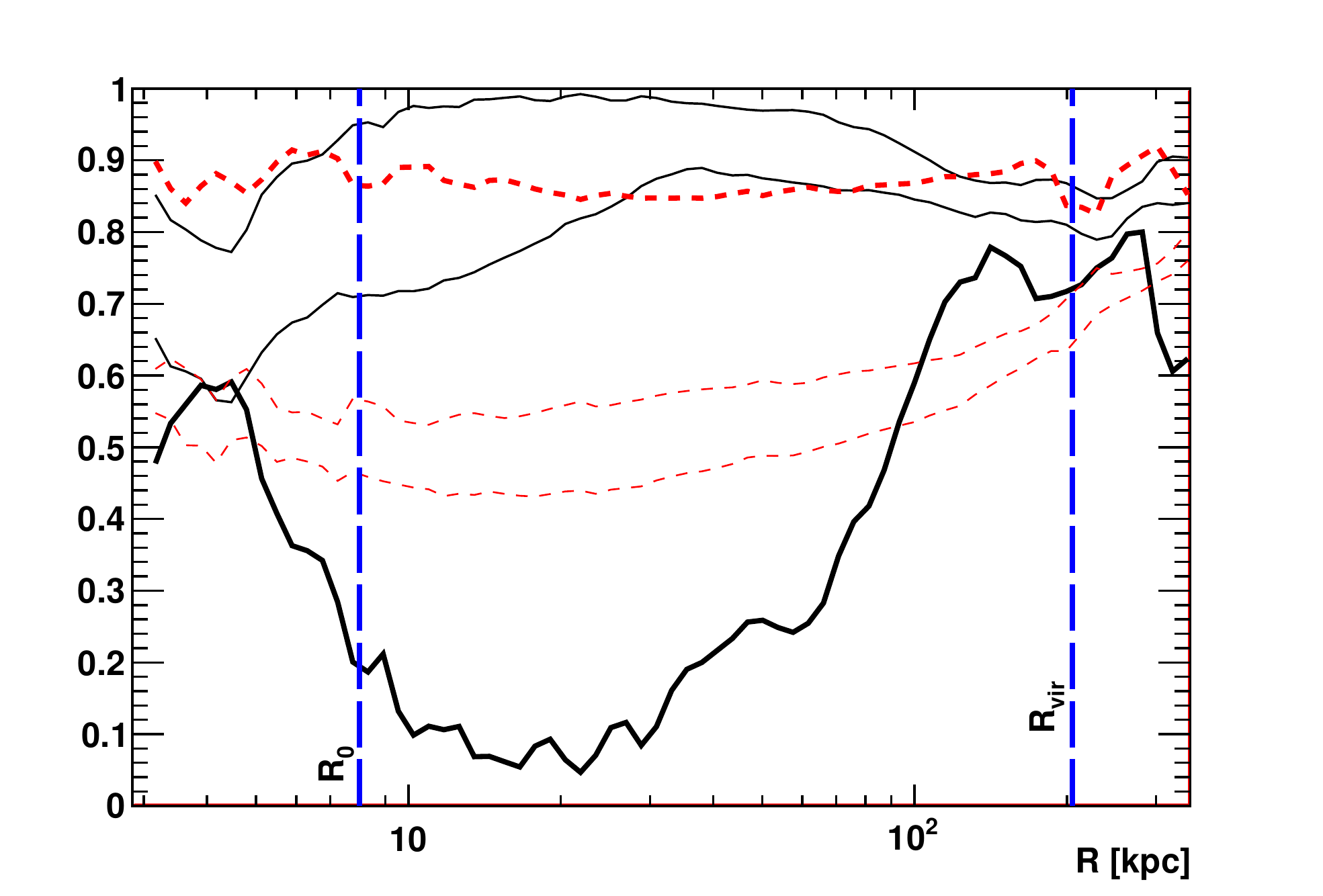}\\
 \caption{\fontsize{9}{9}\selectfont The shape parameters $b/a$, $c/a$ and $T$ for SR6-n01e1ML (solid black lines) and for the corresponding pure dark matter realization (dashed red lines), both at $z=0$. Upper (Lower) thin lines show $b/a$ ($c/a$), whereas thick curves represent the triaxiality parameter $T$. Also shown are the Sun galactocentric distance $R_{0}\simeq 8$ kpc and the virial radius $R_{vir}\equiv R_{200c}$.}\label{figshape1}
\end{figure}

\subsection{Halo shape}
\par It is well-known (e.g. \cite{Gustafsson}) that the inclusion of baryons in numerical simulations washes-out the prolateness of dark matter halos found in DM-only simulations. In order to measure the shape of the dark halo in the simulations under scrutiny (described in section \ref{sims}), we follow \cite{Katz,Debattista} and compute for a given set of $N_p$ dark matter particles the matrix 
\begin{equation}
J_{ij}=\frac{\sum_{k=1}^{N_p}{m_k x_{i,k}x_{j,k}}} {\sum_{k=1}^{N_p}{m_k}}, 
\end{equation}
where $i,j=1,2,3$ index the coordinates in the reference system. The eigenvectors of $J_{ij}$ are the major, intermediate and minor axes $\vec{j}_a$, $\vec{j}_b$ and $\vec{j}_c$, and the eigenvalues $J_a>J_b>J_c$ give the axis ratios through $b/a=\sqrt{J_b/J_a}$ and $c/a=\sqrt{J_c/J_a}$. The major (minor) axis $\vec{j}_a$ ($\vec{j}_c$) corresponds to the axis around which the angular momentum is minimal (maximal). The triaxiality parameter 
\begin{equation*}
T=\frac{1-b^2/a^2} {1-c^2/a^2}
\end{equation*}
distinguishes prolate ($T>0.5$) from oblate ($T<0.5$) shapes. For a given $R$, we start by considering the particles in the sphere of radius $R$ to compute the principle axes, $b/a$ and $c/a$. We then repeat the procedure selecting particles in the ellipsoid $u^2+\frac{v^2}{(b/a)^2}+\frac{w^2}{(c/a)^2}<R^2$, where $u$, $v$ and $w$ are the coordinates along the major, intermediate and minor axes, respectively. The computation is iterated until both $b/a$ and $c/a$ have varied less than 0.5\%.

\par
\begin{figure}[t]
 \centering
 \includegraphics[width=8.5cm,height=6.5cm]{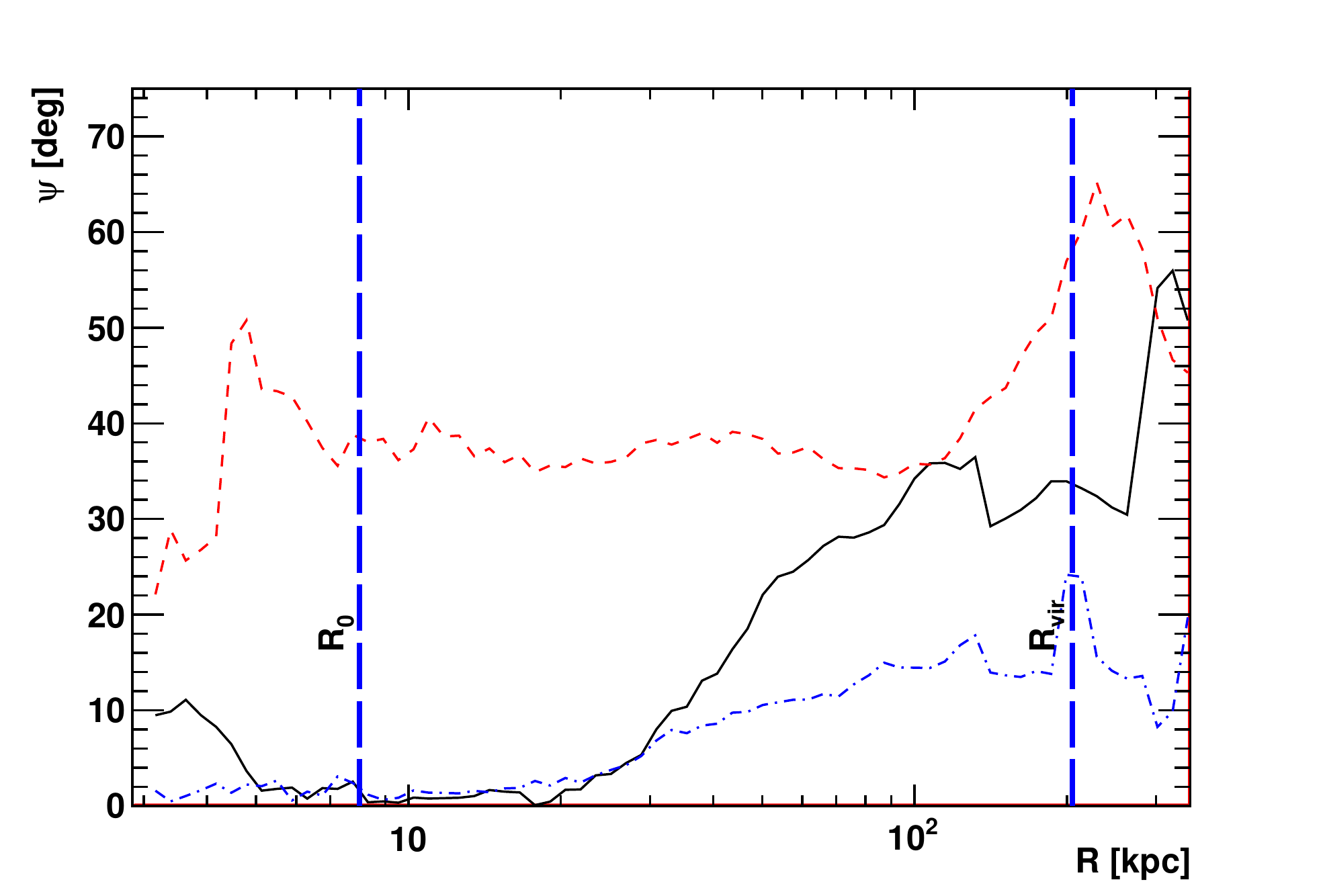}\\
 \caption{\fontsize{9}{9}\selectfont The angle between the normal to the stellar disk $\vec{n}_{sd}$ and the minor axis $\vec{j}_c$. The solid black (dot-dashed blue) line corresponds to SR6-n01e1ML (SR6-n01e5ML) at z=0. The dashed red line shows the angle between the minor axis in the pure dark matter simulation and the normal to the stellar disk in SR6-n01e1ML. Also shown are the Sun galactocentric distance $R_{0}\simeq 8$ kpc and the virial radius $R_{vir}\equiv R_{200c}$.}\label{figshape2}
\end{figure}

 In figure \ref{figshape1} we show $b/a$, $c/a$ (thin lines) and $T$ (thick lines) computed as described in the previous paragraph. Solid and dashed lines correspond respectively to SR6-n01e1ML and the dark matter only realization of the corresponding halo. As expected, in the absence of baryons the dark halo is manifestly prolate i.e.~elongated along the major axis $\vec{j}_a$, while the numerical simulation with baryons produces a more oblate shape i.e.~flattened along the minor axis $\vec{j}_c$. To check the orientation of the dark halo with respect to the baryonic component we plot in figure \ref{figshape2} the angle $\psi$ between $\vec{j}_c$ and the normal to the stellar disk $\vec{n}_{sd}$. In SR6-n01e1ML the dark and baryonic components are fairly aligned for $R<20$ kpc. Furthermore, as the dot-dashed blue line indicates, in SR6-n01e5ML both components are even more aligned. Notice that above $\sim$100 kpc the presence of substructures affects significantly the shape measurement, as clear from figures \ref{figshape1} and \ref{figshape2}.

\begin{figure*}
 \centering
 \includegraphics[width=0.32\textwidth]{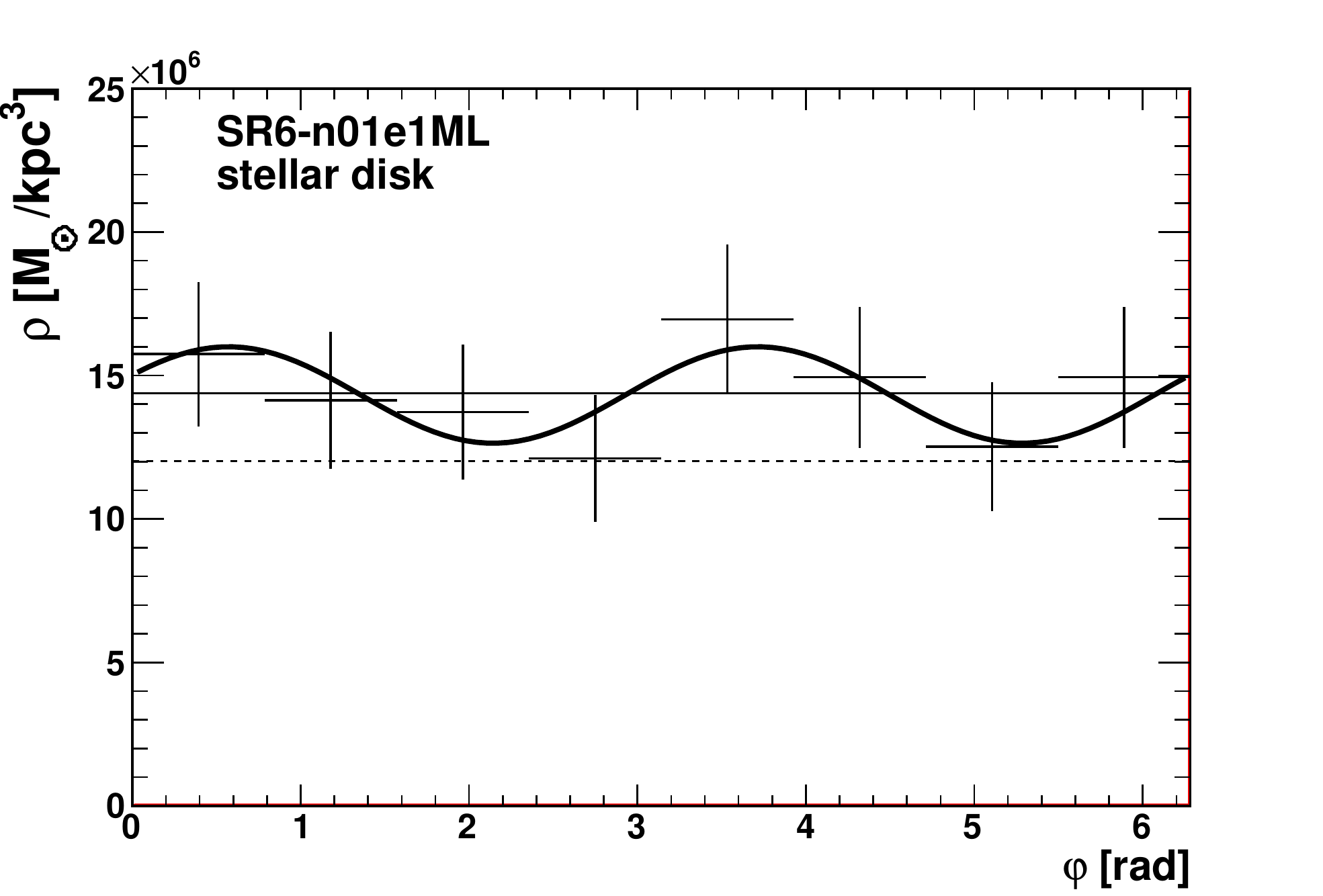}
 \includegraphics[width=0.32\textwidth]{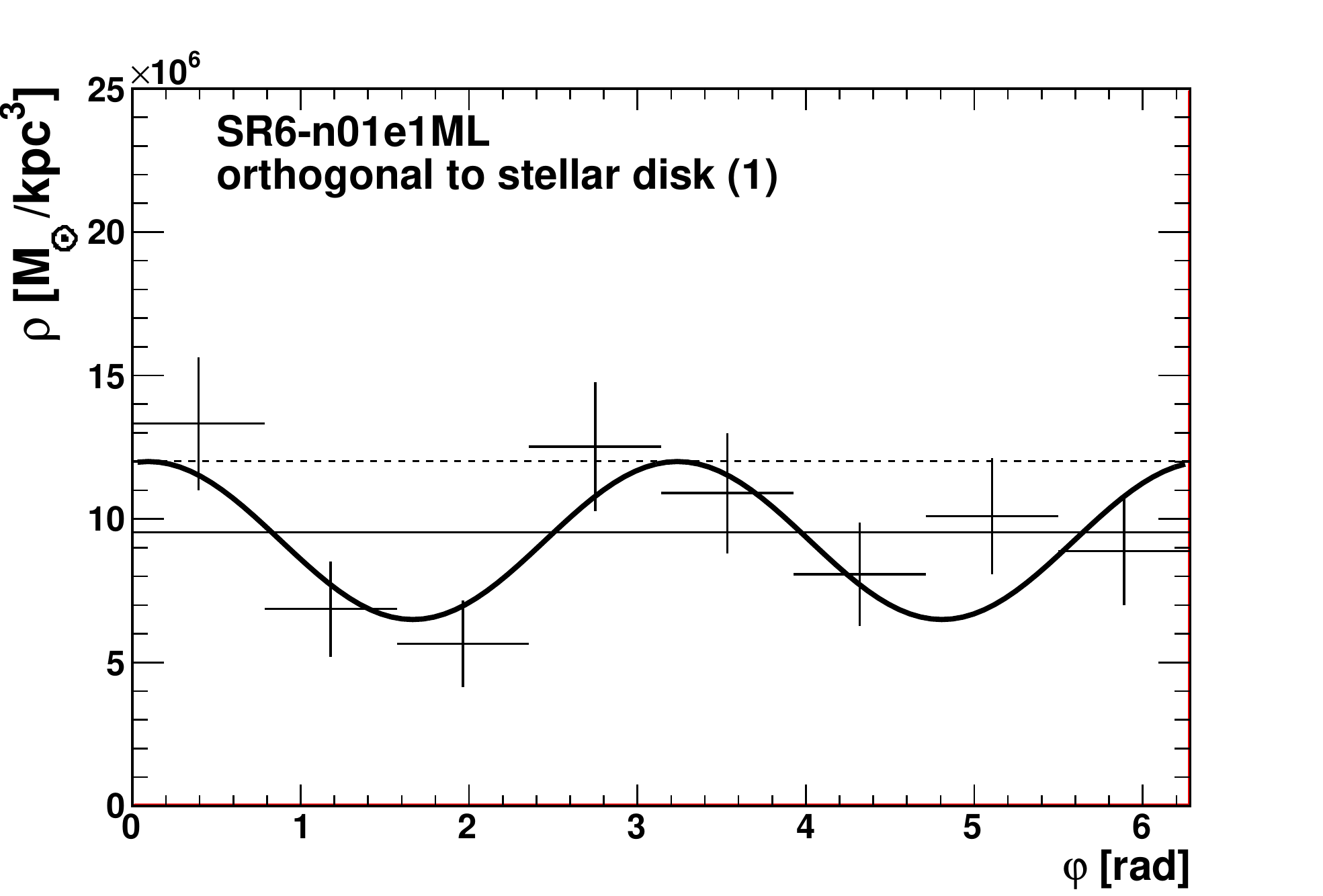}
 \includegraphics[width=0.32\textwidth]{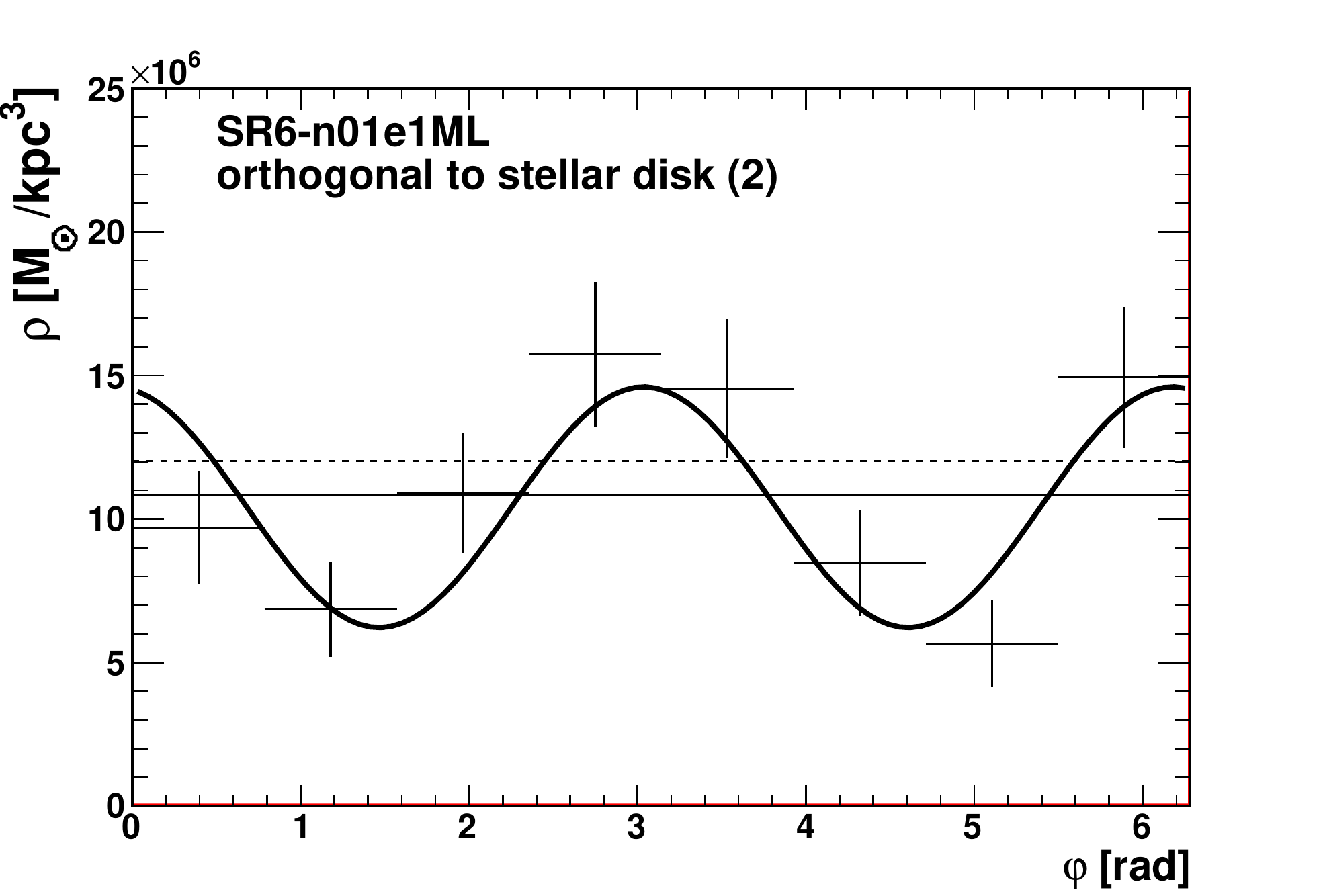}\\
 \includegraphics[width=0.32\textwidth]{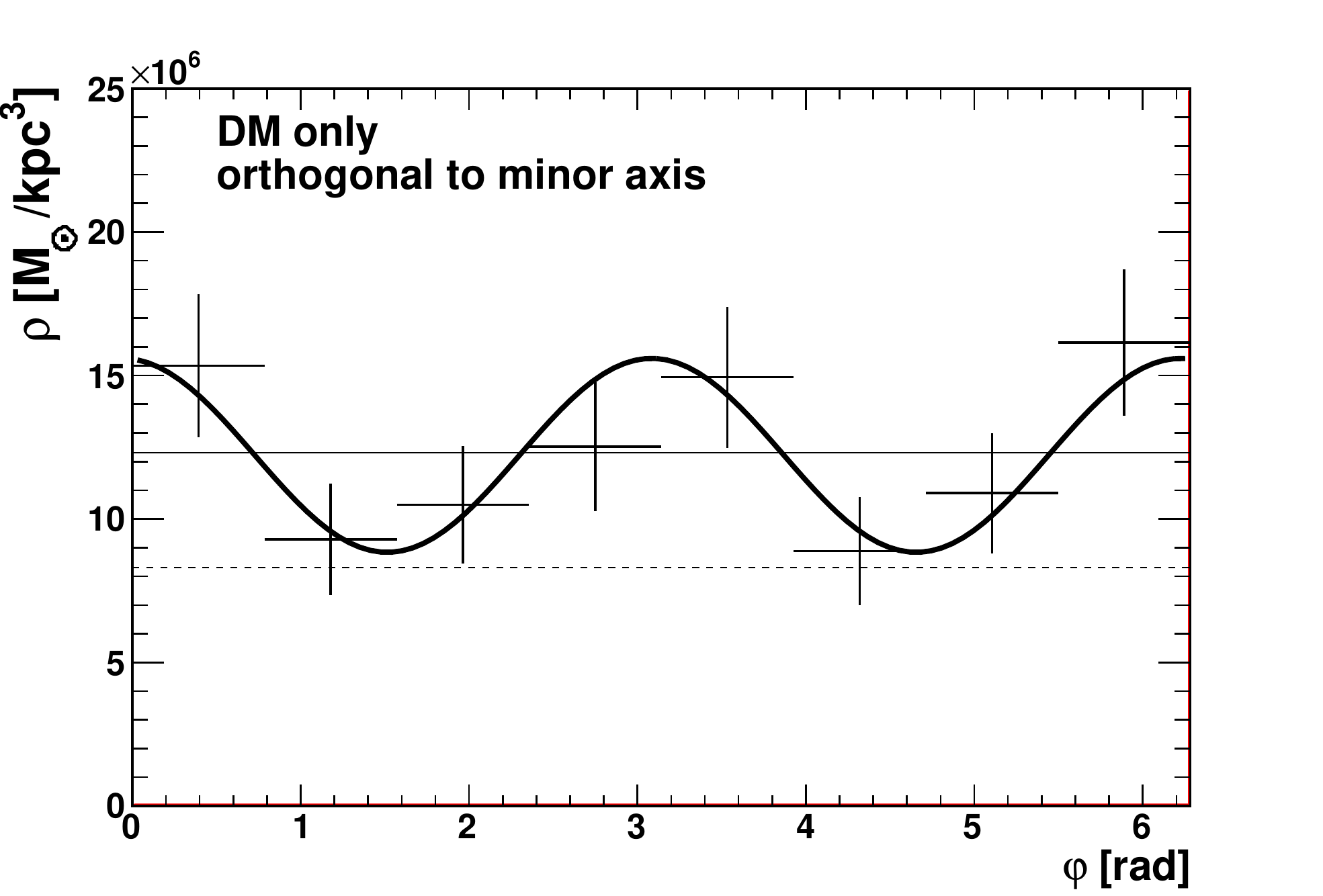}
 \includegraphics[width=0.32\textwidth]{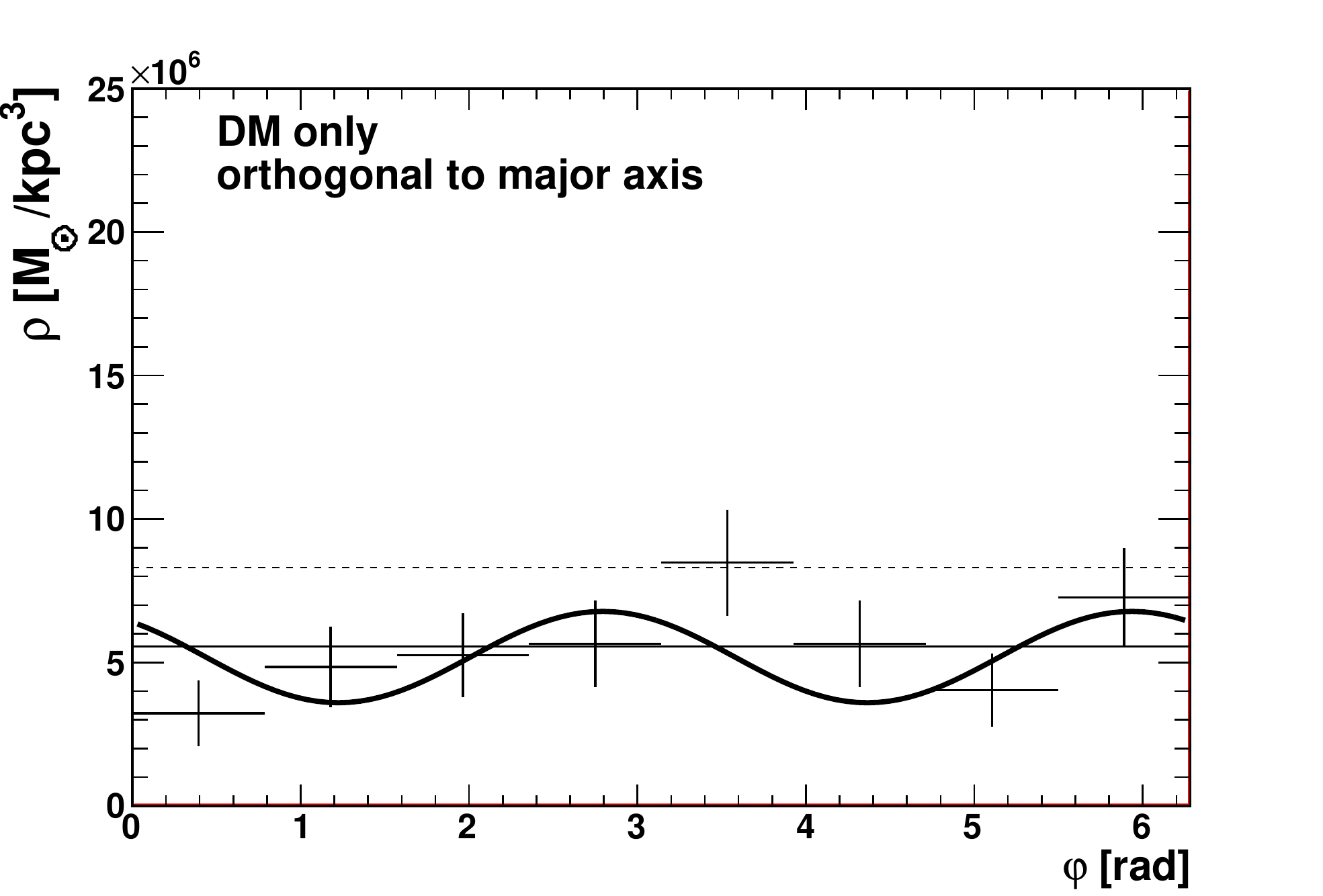}
 \includegraphics[width=0.32\textwidth]{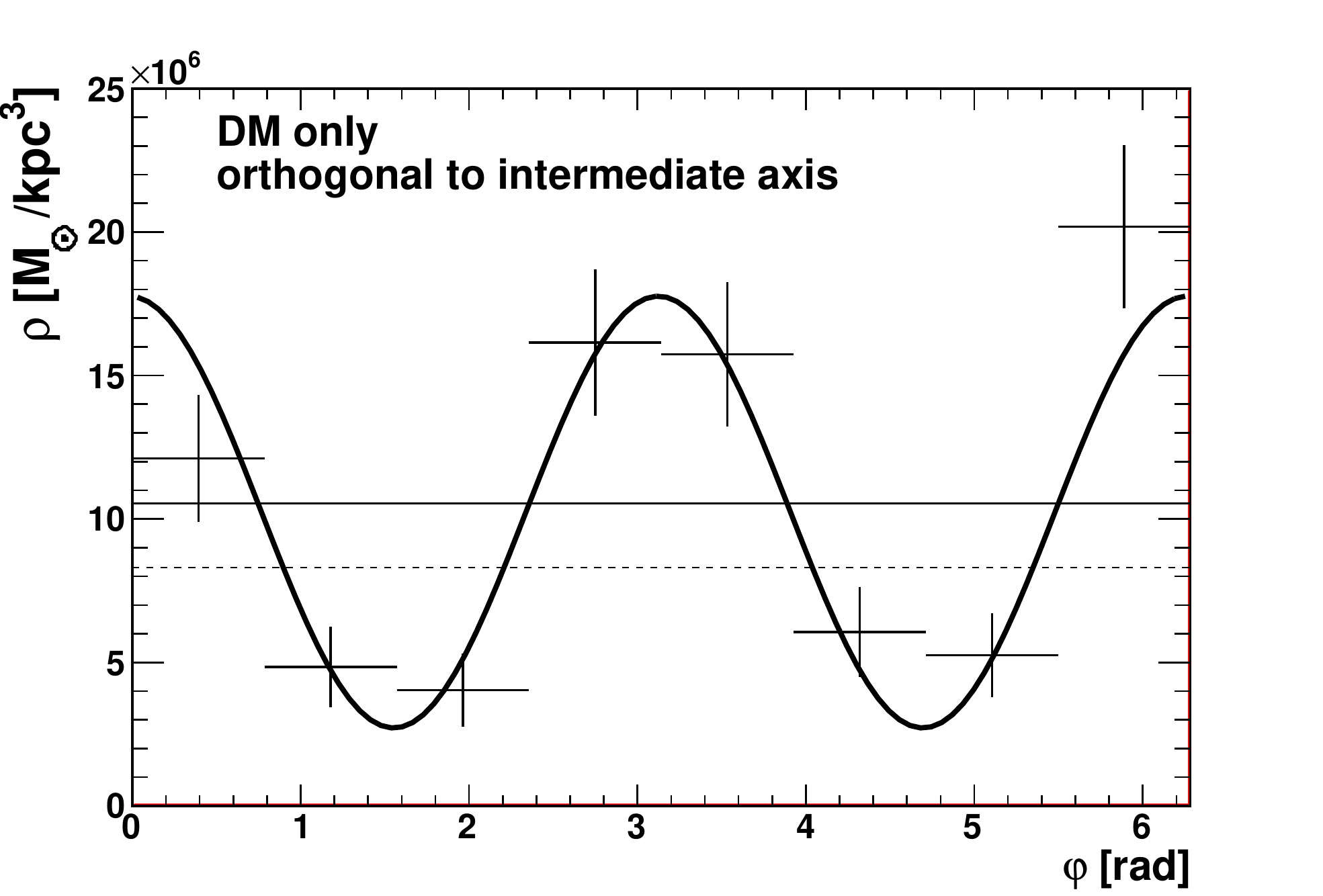}
 \caption{\fontsize{9}{9}\selectfont The dark matter density in the spherical shell $7.5<R/\textrm{kpc}<8.5$ along the stellar disk plane and two perpendicular planes for SR6-n01e1ML (top), and along the planes perpendicular to the principle axes for the pure dark matter simulation (bottom). The solid horizontal line represents the mean of the points and the dashed line shows the value of the mean density in the whole shell, dubbed $\bar{\rho}_0$. The sinusoidal curve shown in each plot is the best fit to the points in the form $c_1+c_2 \, \textrm{sin}\left(2 \, (\varphi + c_3)\right)$.}\label{figprojbaryon}
\end{figure*}

Now, we are interested in evaluating how the determination of $\rho_0$ in studies such as \cite{CatenaUllio} is affected by the DM halo shape, in particular using the latest numerical simulations with baryons. As pointed out in \cite{CatenaUllio}, local observables constrain efficiently $\left.\frac{\partial(v^2 R)}{\partial R}\right|_{R_0}$, where $v$ is the circular velocity in our Galaxy and $R_0$ the Sun galactocentric distance. Such quantity depends on the baryonic content of the Milky Way and on the mass distribution of dark matter through the equations
\begin{equation*}
M_{dm}(<R)\equiv \int{d\varphi\int{d\theta\int_{0}^{R}{dR' \, R'^2 \text{sin}\theta \, \rho(R',\theta,\varphi)}}}
\end{equation*}
\begin{equation}
\frac{1}{G}\left.\frac{\partial(v^2 R)}{\partial R}\right|_{R_0}=K_b+\left.\frac{\partial M_{dm}}{\partial R}\right|_{R_0},
\end{equation}
where $K_b$ encodes the contribution of baryons (at $R_0$ mainly dominated by the disk component). In general,
\begin{equation}
\left.\frac{\partial M_{dm}}{\partial R}\right|_{R_0}=4\pi R_0^2 \bar{\rho}_0,
\end{equation} 
where $\bar{\rho}_0$ is the spherically averaged dark matter density at $R_0\simeq 8$ kpc. We aim at comparing the mean spherical value $\bar{\rho}_0$ with the local one $\rho_0$ in the MW-like simulated galaxy SR6-n01e1ML $-$ such comparison yields the systematic uncertainty in the determination of the local dark matter density presented in works where spherical halos were assumed. In order to study this effect we select dark matter particles that lie inside the spherical shell $7.5<R/\textrm{kpc}<8.5$. Given three orthogonal planes (e.g.~the ones defined by the principal axes) we consider the portions of the shell lying at distances from each plane smaller than $\Delta \omega /2 =0.5$ kpc. This procedure defines three orthogonal ring-like structures each of which we divide in equal parts encompassing an angle $\Delta \varphi = \pi /4$ and thus a volume $V=2\pi$ kpc$^3$. For reference in the following, 
\begin{equation*}
10^7\textrm{ M}_{\odot}/\textrm{kpc}^3 = 0.38\textrm{ GeV/cm}^3.
\end{equation*}
Figure \ref{figprojbaryon} sketches the dark matter density distribution in the above-mentioned rings for SR6-n01e1ML and the dark matter only simulation. In the former case we have used the stellar disk plane and two perpendicular planes, while in the latter the planes defined by the principle axes (of the set of particles in the shell $7.5<R/\textrm{kpc}<8.5$) were considered. In each plot the angle bins represent portions of the ring encompassing $\pi/4$ rad (as described before) and the vertical error bars are Poissonian. The sinusoidal-like modulations seen for both simulations are naturally expected due to the triaxiality of the halos. As a guiding line, we present in each plot of figure \ref{figprojbaryon} the best fit function $c_1+c_2 \, \textrm{sin}\left(2 \, (\varphi + c_3)\right)$. Furthermore, one can appreciate large differences between the spherically averaged density $\bar{\rho}_0$ (dashed lines in figure \ref{figprojbaryon}) and the density along each ring.

\par We pick two extreme cases to bracket the systematic uncertainties in the determination of $\rho_0$: \emph{(i)} the stellar disk plane in SR6-n01e1ML (figure \ref{figprojbaryon}, upper left panel), and \emph{(ii)} the planes perpendicular to the minor and major axes in the dark matter only case (figure \ref{figprojbaryon}, lower left and lower central panels). Notice that we disregard the plane defined by the intermediate axis in the simulation without baryons because in that plane a stable baryonic disk cannot be formed \footnote{We thank J. Diemand for this comment.}. In case \emph{(i)}, since the dark matter halo is flattened along the stellar disk, the local dark matter density is higher than the spherically averaged value:
\begin{equation*}
\rho_0/\bar{\rho}_0=1.01-1.41 \,\, .
\end{equation*}
In case \emph{(ii)} a broader range is obtained:
\begin{equation*} 
\rho_0/\bar{\rho}_0=0.39-1.94\,\, ,
\end{equation*}
in rough agreement with \cite{ghalo}. These values translate into systematic shifts on the local densities found e.g.~in Ref.~\cite{CatenaUllio}.

For the sake of completeness a similar analysis was carried out for the simulated galaxy with efficient star formation rate, SR6-n01e5ML. For the equivalent to case \emph{(i)} explained in the last paragraph, we obtain $\rho_0/\bar{\rho}_0=1.21-1.60$; note nevertheless that this extreme object is not MW-like.

\subsection{Enclosed mass}

\par Both theoretical arguments and simulations of galaxy-sized objects \cite{Blumenthal,Gnedin,Gustafsson} seem to indicate that baryons induce the contraction of the dark matter component towards the central part of the halo. This is also the case for the simulated galaxy SR6-n01e1ML when compared to the corresponding pure dark matter realization. Note that the dark matter only simulation needs to be rescaled down by a factor $(\Omega_m-\Omega_b)/\Omega_m\simeq 0.8333$ in order to account for the presence of baryons, and that one can only firmly trust the numerical simulation results above about 2 times the resolution scale, i.e.~680 pc. We leave for a forthcoming work the detailed study of the DM profile and adiabatic contraction models. This topic, addressed in Refs.~\cite{Pedrosa,Abadi,Gustafsson,Tissera} in the framework of different numerical simulations, is of particular importance for indirect dark matter searches.


\par A relevant quantity to analyse when trying to determine the local dark matter density is the enclosed dark mass, $M_{dm}(<R)$, since this affects the rotation curve of the Galaxy. Using the spherically averaged density reported in figure \ref{figprojbaryon} (dashed lines), we find that the same enclosed mass $M_{dm}(<8\textrm{ kpc})$ would result in SR6-n01e1ML and the pure dark matter halo if the local DM densities are rescaled such that $\bar{\rho}_0(\textrm{SR6-n01e1ML})/\bar{\rho}_0(\textrm{DM only})\simeq 0.9$. The lower density in the presence of baryons is simply a reflex of a more concentrated profile. In any case, these estimates do not translate directly into systematic uncertainties in the determination of $\rho_0$ since precise determinations of local observables $-$ namely the Oort's constants $A\pm B$, the Sun galactocentric distance $R_0$ and the local visible matter surface dentity $\Sigma_{\ast}$, see \cite{CatenaUllio,Salucci} $-$ constrain efficiently both $\bar{\rho}_0$ (through $\left.\frac{\partial(v^2 R)}{\partial R}\right|_{R_0}$) and $M_{dm}(<R_0)$ (through $v(R_0)$). Therefore, we conclude that considering a contracted dark matter profile would not change significantly the determination of the local dark matter density from precise dynamical observables, but would eventually prefer smaller concentration parameters (or, equivalently, larger scale radii $R_s$).

\section{Conclusions}\label{conc}
\par The dark matter density in our neighbourhood is the key astrophysical ingredient that fixes the flux of DM particles crossing the Earth and the Sun, thus governing the scattering off nuclei in underground detectors as well as the capture rate in the Sun. Experiments looking for DM-induced nuclei recoils or neutrino fluxes from the Sun are hence crucially dependent on the local dark matter density. In the present work we have tried to quantify the systematic uncertainties associated to this parameter, and that affect determinations based on dynamical observables of our Galaxy. Using a very recent successful attempt to simulate a spiral galaxy that resembles the Milky Way, the dark matter density at the solar circle was analysed in detail and compared to the pure dark matter case. 

One major consequence of the inclusion of baryons is a significant flattening of the dark halo in the direction of the normal to the stellar disk, leading to a DM overdensity in the local disk of up to 41\% with respect to the spherically averaged value. More specifically, we found that in the MW-like simulated galaxy the local dark matter density is higher than the spherically averaged value: $\rho_0/\bar{\rho}_0=1.01-1.41$. In the DM-only case a broader range is obtained: $\rho_0/\bar{\rho}_0=0.39-1.94$. 

Ideally, one should repeat the analysis in  Ref.~\cite{CatenaUllio}, i.e. a Bayesian approach to mass modeling of the Milky Way components, in presence of a triaxial profile like the one discussed here. However, based on the considerations presented above, a better estimate of the local dark matter density can be obtained by raising by 21\% the mean value obtained in Ref.~\cite{CatenaUllio} for the spherical case, keeping relative statistical errors fixed and adding systematic errors. In the case of an Einasto profile, this procedure suggests
\begin{equation}
\rho_0=0.466\pm0.033 {\textrm{(stat)}} \pm0.077 {\rm (syst)} {\textrm{ GeV}}/ \textrm{cm}^{3} \,\, .
\nonumber
\end{equation}
Notice that the mean 21\% enhancement with respect to the spherical local DM density is obtained for a specific simulated galaxy resembling the MW. The actual enhancement in our Galaxy may, of course, be different, but the main points here are that \emph{(i)} the presence of baryons leads quite generally to a DM overdensity about the local disk, and \emph{(ii)} the systematic uncertainties affecting $\rho_0$ are significant and, in some cases, already larger than the statistical ones.

\par The baryons are also responsible for a non-negligible contraction of the DM distribution towards the central part of the galaxy. Even though this may be very important in searching for products of DM annihilations $-$ such as positrons or antiprotons from the galactic halo, and $\gamma$-rays or neutrinos from the Galactic Centre $-$ we found that it has no significant effect in the determination of the local dark matter density using dynamical observables.

\par An estimate of systematic uncertainties affecting the local dark matter density is an important step in assessing realistically our present knowledge on this key parameter. Such knowledge is in turn an input in interpreting direct detection results, combining multiple DM-induced signals, and extracting compatible DM properties. Another relevant ingredient for DM scattering and capture is the phase space density (see e.g.~\cite{kuhlen,mccabe}), namely the distribution of velocities $f(|v|)$ $-$ such topic is out of the scope of this work, but it would be interesting to study it in the set of simulations analysed here. In case of a positive signal in direct detection experiments, for instance, the identification of the compatible particle physics parameter space and the discrimination between different particle physics frameworks depend crucially upon the state of our knowledge in key astrophysical parameters such as the dark matter density in our neighbourhood.

{\it Acknowledgements:} We thank J\"urg Diemand for useful conversations. MP is supported by Funda\c{c}\~{a}o para a Ci\^encia e Tecnologia (Minist\'erio da Ci\^encia, Tecnologia e Ensino Superior).

\end{document}